\documentclass[reprint,amsmath,amssymb,aps,twocolumn,showpacs,
superscriptaddress,nofootinbib]{revtex4-1}

\usepackage{graphicx}
\usepackage{epsfig}
\usepackage[hang]{subfigure}
\usepackage{epstopdf}
\usepackage{color}
\usepackage[normalem]{ulem}

\usepackage{hyperref}

\usepackage[
]{todonotes}

\newcommand{\be}{\begin{equation}}
\newcommand{\ee}{\end{equation}}
\newcommand{\ba}{\begin{eqnarray}}
\newcommand{\ea}{\end{eqnarray}}
\newcommand{\beal}{\begin{aligned}}
\newcommand{\eeal}{\end{aligned}}
\newcommand{\beqa}{\begin{eqnarray}}
\newcommand{\eeqa}{\end{eqnarray}}
\newcommand{\nn}{\nonumber}

\begin{document}

\title{Holographic Thermodynamics of Accelerating Black Holes}
\author{Andr\'{e}s Anabal\'{o}n}
\affiliation{Universidad Adolfo Ib\'{a}\~{n}ez; Dep.\ de Ciencias,
Facultad de Artes Liberales, Av. Padre Hurtado 750, - Vi\~{n}a del Mar, Chile.}
\author{Michael Appels}
\affiliation{Centre for Particle Theory, Durham University, South Road,
Durham, DH1 3LE, UK}
\author{Ruth Gregory}
\affiliation{Centre for Particle Theory, Durham University, South Road,
Durham, DH1 3LE, UK}
\affiliation{Perimeter Institute, 31 Caroline St., Waterloo, Ontario,
N2L 2Y5, Canada}
\author{David Kubiz\v n\'ak}
\affiliation{Perimeter Institute, 31 Caroline St., Waterloo, Ontario,
N2L 2Y5, Canada}
\author{Robert B. Mann}
\affiliation{Department of Physics and Astronomy, University of Waterloo, Waterloo,
Ontario, Canada, N2L 3G1}
\affiliation{Perimeter Institute, 31 Caroline St., Waterloo, Ontario,
N2L 2Y5, Canada}
\author{Ali \"{O}vg\"{u}n}
\affiliation{Instituto de F\'{\i}sica, Pontificia Universidad Cat\'olica de
Valpara\'{\i}so, Casilla 4950, Valpara\'{\i}so, Chile}
\affiliation{Physics Department, Arts and Sciences Faculty, Eastern Mediterranean
University, Famagusta, 99628, North Cyprus via Mersin 10, Turkey}

\date{November 11, 2018}

\begin{abstract}
We present a careful study of accelerating black holes in anti-de Sitter
spacetime, formulating the thermodynamics and resolving discrepancies
that have appeared in previous investigations of the topic. We compute
the dual stress-energy tensor for the spacetime and  identify the energy
density associated with a static observer at infinity. The dual energy-momentum
tensor can be written as a three-dimensional perfect fluid plus a non-hydrodynamic
contribution with a universal coefficient which is given in gauge theory variables.
We demonstrate that
both the holographic computation and the method of conformal completion
yield the same result for the mass. We compare to previous work on
black funnels and droplets, showing that the boundary region can be
endowed with non-compact geometry, and comment on this novel holographic
dual geometry.
\end{abstract}

\pacs{04.20.Jb, 04.70.-s}
\maketitle

The importance of black holes in advancing our understanding of physics cannot
be underestimated.  They provide a setting for testing our most fundamental
ideas about gravity under extreme conditions and offer us insight into the
underlying microscopic degrees of freedom that may associated with quantum
gravity.  The subject of black hole thermodynamics~\cite{Bekenstein:1973ur,
Bekenstein:1974ax,Hawking:1974sw} has proven to be an invaluable tool to this
end, and broad classes of black holes have been shown to exhibit a rich and
varied range of thermodynamic behaviour, particularly in anti-de Sitter
spacetime~\cite{Kubiznak:2016qmn}.

Within this framework, accelerating black holes have presented a
unique challenge.  The idealized solution is described by the
\emph{C-metric}~\cite{Kinnersley:1970zw,
Plebanski:1976gy,Dias:2002mi,Griffiths:2005qp}, whose spacetime has a
string-like singularity along one polar axis attached to the black hole.
We can think of this conical singularity as a cosmic string (indeed, the
conical singularity can be replaced by a finite width topological
defect~\cite{Gregory:1995hd}, or a magnetic flux tube~\cite{Dowker:1993bt})
with the tension providing the force driving the acceleration.
Surprisingly, even though these black holes are not isolated because of
the ``cosmic strings'' it is possible to derive sensible looking thermodynamics,
although recent studies have apparently conflicting results
\cite{Appels:2016uha,Appels:2017xoe,Gregory:2017ogk,Astorino:2016ybm}.

We consider here the interpretation of an accelerating black hole in
anti-de Sitter (AdS) spacetime, with a focus on a holographic
interpretation of the thermodynamics. We resolve  conflicting issues that
exist in the literature, obtain a distinct set of thermodynamic
variables that are now consistent with the gravitational action,
and agree with both the conformal and holographic
methods for computing conserved charges. To this end, we focus
our attention to black holes with no acceleration horizon
\cite{Podolsky:2002nk}, so that there is no ambiguity as to which
horizon temperature should be considered, or as to whether there is
an equilibrium thermodynamics for the system. In addition, as we discuss,
the holographic computation and interpretation are also unambiguous
and straightforward. We also comment on the cases when the acceleration
horizons appear and provide a novel interpretation of the boundary geometry.

\begin{figure}[tbp]
\centering
\includegraphics[width=0.45\textwidth]{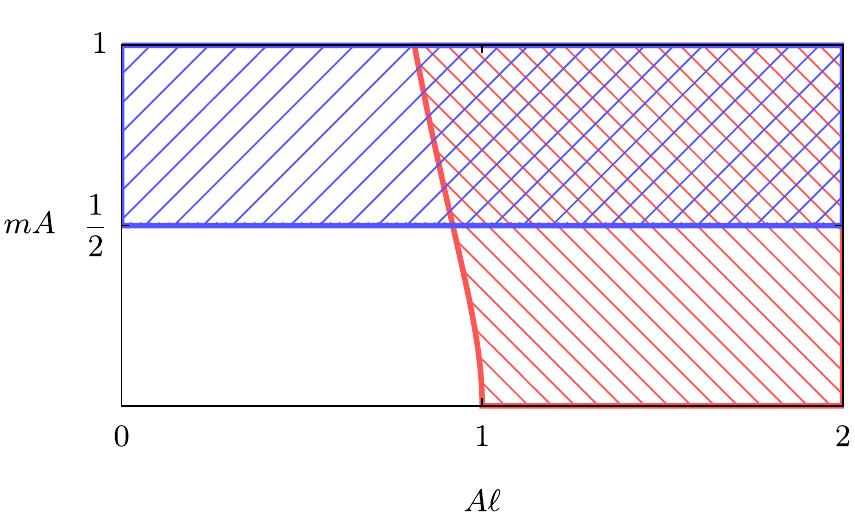}
\caption{
\textbf{Parameter space.}
The blue and red lines denote the boundaries in the parameter space
$(mA,A\ell)$ for which the holographic computation is valid.
The hashed red region is where acceleration horizons are present and
the hashed blue region is where the metric signature is not preserved,
leaving the white region as the physical parameter space.}
\label{fig:f1}
\end{figure}

An accelerating black hole in AdS can be described by the metric
\cite{Podolsky:2002nk, Hong:2003gx, Griffiths:2005qp}
\begin{equation}\label{AdSC}
ds^2=\frac{1}{\Omega^2}\bigg[ -fdt^2+\frac{dr^2}{f}
+r^2\Big(\frac{d\theta^2}{\Sigma}
+\Sigma\sin^2\!\theta\frac{d\phi^2}{K^2}\Big)\bigg]\,,
\end{equation}
where
\be
\beal
\Omega&=1+Ar\cos\theta\,, \qquad  \Sigma= 1+2mA\cos\theta\,, \\
f(r)&=(1-A^2r^2)\bigg(1-\frac{2m}{r}\bigg)+\frac{r^2}{\ell^2}\,.
\eeal
\ee
The potential $f(r)$ shows clearly the black hole nature of the solution,
as well as the effects of acceleration ($A$) and cosmological constant
{($\Lambda= -3/\ell^2$)}.
Note that we require $2mA<1$ to preserve the
metric signature.  The parameter $K$ encodes
information about the conical deficits on the north and south
poles that have tensions given by~\cite{Appels:2017xoe}
\begin{equation}
\label{mueq}
\mu_{\pm}=\frac{\delta_\pm}{8\pi}
=\frac{1}{4}\Big(1-\frac{\Sigma(\theta_\pm)}{K}%
\Big)=\frac{1}{4}\Big(1-\frac{1\pm 2mA}{K}\Big)\,.
\end{equation}
The absence of an acceleration horizon yields the
constraint $f(-1/A\cos\theta) > 0$, in turn constraining the parameter space
$(m,\ell)$ to the white region bounded by the blue and red lines in figure
\ref{fig:f1}.  It is straightforward to show via a linear transformation
\cite{Hong:2003gx} on the coordinates $(x=\cos\theta,y=-1/Ar)$ that the latter
bound is equivalent to the absence of black droplets \cite{Hubeny:2009kz}.

As discussed in \cite{Appels:2017xoe,Gregory:2017ogk}, setting $m=0$
removes the black hole horizon, and leaves pure  AdS spacetime in Rindler-type
coordinates.  Performing the coordinate transformation~\cite{Podolsky:2002nk}:
\begin{equation}
1+\frac{R^2}{\ell^2}=\frac{1+(1-A^2\ell^2)r^2/\ell^2}{(1-A^2\ell^2)\Omega^2}\,, \quad
R\sin\vartheta=\frac{r\sin\theta}{\Omega}\,,
\end{equation}
recovers AdS in global coordinates:
\begin{equation}
\label{gAdS}
 ds^2_{AdS}= -\Big(1+\frac{R^2}{\ell^2}\Big) \alpha^2
dt^2+\frac{dR^2}{1+\frac{R^2}{\ell^2}}  +R^2\Big(d\vartheta^2+\sin^2\vartheta
\frac{d\phi^2}{K^2}\Big)\,,
\end{equation}
but with the notable feature that the time coordinate is not the expected AdS time, but is
rescaled by a factor of $\alpha = \sqrt{1-A^2 \ell^2}$. Conventionally, we
choose the normalisation of our time coordinate so that it corresponds to
the ``time'' of an asymptotic observer. While this is potentially a slightly slippery
concept in AdS, taken together with the spherical asymptotic spatial coordinates,
this scaling suggests that the correct time coordinate is not in fact $t$, but rather
$\tau=\alpha t$, giving a rescaling of the time-coordinate in \eqref{AdSC}.
If we now proceed with this metric, and compute the temperature
associated with the black hole (also the temperature of the boundary
field theory), then we obtain
\be
T =\frac{f'(r_+)}{4\pi\alpha}
=\frac{1 + 3\frac{r_+^2}{\ell^2}
- A^2r_+^2 \left (2+\frac{r_+^2}{\ell^2}-A^2r_+^2\right)}
{4\pi \alpha r_+(1-A^2r_+^2)}\,,
\label{temp}
\ee
where $f(r_+)=0$.

It is worth pausing to reflect on this result. In past work
\cite{Appels:2016uha,Appels:2017xoe,Gregory:2017ogk},
the standard time coordinate appearing in the AdS C-metric was used
to derive the temperature of the black hole horizon.
This appeared to
be a natural approach as the blackening factor of the metric was in its
canonical form. However, as pointed out in \cite{Gibbons:2004ai}, normalising the
time and timelike Killing vector is key to obtaining the correct
thermodynamics, although the method of obtaining this correct
normalisation was less transparent. Here, having uncovered this
suggestive result, we now proceed carefully with considering
thermodynamics of the accelerating black hole.
As usual, we will take the entropy to be one quarter of the horizon area:
\be
S=\frac{\mathcal{A}}{4}=\frac{\pi r_+^2}{K(1-A^2r_+^2)}\,.
\label{entropy}
\ee
The remaining task is to correctly identify the black hole mass, often the
biggest challenge in studying thermodynamics of black holes with non-trivial
asymptotics. In what follows, we will provide two independent arguments,
beginning with the conformal completion method \cite{Ashtekar:1999jx,Das:2000cu}.  Although consistency
of the thermodynamic relations is a common method for deriving
thermodynamics (used for {example in \cite{Astorino:2016ybm}}), we do not
consider this sufficient;  hence we return to our theme of holography,
computing the holographic stress tensor of the boundary
theory, thereby confirming our result.
As an ancillary argument, we finally check consistency with
a computation of the free energy.

The first argument uses the Ashtekar--Das definition of conformal
mass \cite{Ashtekar:1999jx,Das:2000cu}, which extracts the mass
via conformal regularisation of the AdS C-metric near the boundary.
The idea is to perform a conformal transformation on \eqref{AdSC},
$\bar{g}_{\mu\nu} = {\bar{\Omega}}^{2}g_{\mu\nu}$,
to remove the divergence near the boundary, then  obtain a
conserved charge by  integrating the conserved current
 \be
Q(\xi )=\frac{\ell}{8\pi}\lim_{\bar{\Omega} \rightarrow 0}\oint
\frac{\ell^{2}}{\bar{\Omega}}N^{\alpha }N^{\beta }
\bar{C}^{\nu}{}_{\alpha \mu \beta }
\xi _{\nu }d\bar{S}^{\mu }
\ee
composed of the Weyl tensor of the conformal metric,
$\bar{C}^{\mu }{}_{\alpha \nu \beta }$, the normal
to the boundary, $N_{\mu }=\partial _{\mu }{\bar{\Omega}}$, and a
suitable Killing vector for the mass, $\xi = \partial_\tau$.
Even though the conformal completion is not unique, the
charge thus obtained is independent of the choice of conformal completion.
We pick $\bar{\Omega}=\ell\Omega r^{-1} $, which provides a smooth
conformal completion in the limit $A=0$. The spacelike surface element
tangent to $ \bar{\Omega}=0$ is
\begin{equation}\label{surfel}
d\bar{S}_{\mu }=\delta^{\tau }_{\mu }\frac{\ell^{2} (d\cos\theta) d\phi}{\alpha K}\,,
\end{equation}%
obtained by inserting $Ar\cos\theta=-1$ into the metric  $\bar{g}_{\mu\nu}$
and computing the relevant determinant.  This yields
\begin{equation}\label{confmass}
M = Q(\partial _{\tau})=\alpha \frac{m}{K}
\end{equation}
for the mass, in agreement with the  temperature \eqref{temp}, but in contrast
to previous results~\cite{Appels:2016uha,Astorino:2016ybm}.
The absence of acceleration horizons ensures that  $M$ vanishes in the
limit $A\ell\to 1$ only for $m=0$ and is positive otherwise.

It is now straightforward to verify the first law and Smarr \cite{Smarr:1972kt}
relation
\begin{align}
\delta M &=T\delta S+V\delta P-\lambda _{+}\delta \mu _{+}
-\lambda_{-}\delta \mu _{-}\,,  \nn \\
M &=2TS-2PV\,,
\end{align}
using \eqref{temp}, \eqref{entropy}, and \eqref{confmass},
provided
\be
\beal \label{vol}
V &= \frac{4}{3}\frac{\pi}{K \alpha}\left[\frac{r_+^3}{(1-A^2r_+^2)^2}
+mA^2\ell^4\right]\,, \\
\lambda_\pm &= \frac{1}{\alpha}\left[\frac{r_+}{1-A^2r_+^2}
-m\left(1\pm\frac{2A\ell^2}{r_+}\right)\right]\,,
\eeal
\ee
where $P = 3/8\pi \ell^2$ is the thermodynamic pressure associated with
the cosmological constant  \cite{Kubiznak:2016qmn}, and $\lambda_\pm$
are the thermodynamic lengths introduced in \cite{Appels:2017xoe,Gregory:2017ogk}
{that are conjugate to the tensions.}
We have included the possibility that the tensions vary, as otherwise
the system is constrained and identification of the correct parameters can be
misleading.

We now turn to another method for deriving the thermodynamic mass,
by computing the holographic stress tensor. This provides an alternate
and completely independent method of computation, and will reveal
the dual interpretation of this system.  The idea here is to perform a
Fefferman--Graham expansion of the metric \cite{FG}, identifying the
fall-off of sub-leading terms in the metric at the boundary. These are
then used to compute the dual stress-energy tensor that can be integrated
to give the mass of the system.

The action, including boundary counterterms
\cite{Balasubramanian:1999re, Mann:1999pc}, is
\be
\beal
I[g]=&\frac{1}{16\pi}\int_{M}d^{4}x\sqrt{-g}\left[ R+\frac{6}{\ell^{2}}\right]
+ \frac{1}{8\pi}\int_{\partial M}d^{3}x\sqrt{-h}\mathcal{K} \\
&- \frac{1}{8\pi}\int_{\partial M}d^{3}x\sqrt{-h}
\left[ \frac{2}{\ell}+\frac{\ell}{2}\mathcal{R}\left( h\right) \right]\,,
\eeal
\label{action}
\ee
where $\mathcal{K}_{ab}$ is the extrinsic curvature of the boundary
metric, evaluated asymptotically in an appropriate coordinate system,
defined presently. $h_{ab}$ is the intrinsic metric on
$\partial {\cal M}$, and ${\cal R}$ its Ricci curvature. Varying
the action gives the energy momentum tensor:
\begin{equation}
8\pi \mathcal{T}_{ab} = \ell \mathcal{G}_{ab}\left(h\right)
-\frac{2}{\ell}h_{ab}-\mathcal{K}_{ab}+h_{ab}\mathcal{K}\,.
\end{equation}%

To compute these terms requires new coordinates near the boundary
of AdS, typically parametrised by Fefferman--Graham
coordinates, in which
\be
ds^2= \frac{\ell^2}{\rho^2} d\rho^2 +
\frac{\rho^2}{\ell^2} \left ( \gamma_{ab}^{(0)} + \frac1{\rho^2} \gamma^{(2)}_{ab}
+ ...\right )  dx^a dx^b\,.
\label{FGmetric}
\ee
Although often one identifies a $\rho$ coordinate globally, due to the
complexity of \eqref{AdSC}, we instead perform an asymptotic
expansion for the coordinate transformation, writing
\begin{equation}\label{FGt}
\frac{1}{Ar}= -x-\sum F_{n}\left( x\right) \rho ^{-n}\,, \quad
\cos \theta = x+\sum G_{n}\left( x\right) \rho ^{-n}\, .
\end{equation}
The functions $F_n$ and $G_n$ are fixed by the required fall-off properties
of \eqref{FGmetric}, apart from $F_1$, that we choose to write as
\begin{equation}\label{F1}
F_{1}(x) = - \frac{\left( 1-A^{2}\ell^{2}X\right) ^{3/2}}{A\omega(x) \alpha}\,,
\end{equation}
in order to elucidate the conformal degree of freedom in the boundary metric,
$\omega$, with $ X = (1-x^{2})\left( 1+2mAx\right)$.
Computing this boundary metric, $ds_{(0)}^2 = \gamma^{(0)}_{ab} dx^a dx^b$,
we find it sufficient to truncate the series \eqref{FGt} at $n=4$ and find:
\be\label{bndymet}
{ds_{(0)}^2}
= -\omega^{2}d\tau^2
+\frac{\omega^2\alpha^2\ell^{2} dx^2}{X(1-A^2\ell^2X)^2}
+ \frac{X\omega^2\alpha^2 \ell^{2}d\phi^2}{K^2(1-A^2\ell^2X)}\,.
\ee
Note that the  transformation \eqref{FGt} is valid in general only when
$A^{2}\ell^{2}X<1$, which is precisely the constraint that acceleration
horizons are absent.

The expectation value of the energy momentum of the CFT$_{3}$
can then be calculated, yielding
\begin{equation}
\left\langle \mathcal{T}_{a}^{b}\right\rangle =\lim_{\rho \longrightarrow
\infty }\frac{\rho }{\ell}\mathcal{T}_{a}^{b}=\left(
\begin{array}
[c]{ccc}%
-\rho_{E} & 0 & 0\\
0 & \frac{\rho_{E}}{2}+\Pi & 0\\
0 & 0 & \frac{\rho_{E}}{2}-\Pi
\end{array}
\right)\,,  \label{EMT}%
\end{equation}
where
\begin{equation}
\beal
\rho _{E} & = \frac{m}{8\pi\ell^2 \alpha^3\omega^3}(1-A^2\ell^2 X)^{3/2}
(2-3A^{2}\ell^{2}X)\,,\\
\Pi &= \frac{3mA^{2}\textcolor{red}{}X}{16\pi \alpha^3\omega^{3}}\left(
1-A^{2}\ell^{2}X\right) ^{\frac32} \,.
\eeal
\end{equation}
We can re-express this in the language of the fluid/gravity correspondence (for a review and references see
\cite{Rangamani:2009xk}) as
\begin{equation}
\left\langle \mathcal{T}_{ab}\right\rangle = \frac32 d(x)
U_{a}U_{b}+\frac{d(x)}2 \gamma^{(0)} _{ab}+\xi\Theta _{ab}\,,  \label{T}
\end{equation}
where the
4-velocity $U$ and $\Theta$ are defined as
\be
\beal
U&=\sqrt{\frac{3-2A^2\ell^2 X}{3\omega (1-A^2\ell^2 X)}}\partial_\tau
- \frac{A X}{\sqrt{3\omega}\alpha} \sqrt{(1-A^2\ell^2 X)} \partial_x\,,\\
\Theta_{ab} &=C_{abd}U^{d}+C_{bad}U^{d}\,,
\eeal
\ee
with $C_{abc} = \nabla_{[c} R_{b]a} - g_{a[b}\nabla_{c]}R/4$  the
Cotton tensor \cite{Mukhopadhyay:2013gja} {of the metric \eqref{bndymet}};
boundary indices are raised and lowered with $\gamma^{(0)}_{ab}$.

The fact that the boundary is non-conformally flat leads to the inclusion
of non-hydrodynamic corrections to the perfect fluid, with
\be
d(x) = \frac{m \sqrt{(1-A^2\ell^2 X)^5} }{4\pi  \ell^2 \omega^3 \alpha^3}\,.
\ee
The universal (in the sense of independent of black hole mass
and acceleration) coefficient is
\be \xi=\frac{\ell^{2}}{8\pi \sqrt{3}} =\sqrt{\frac{2}{3}}\frac{1}{12\pi} k^{1/2}N^{3/2}\,,
\ee
using the standard
$AdS_4/CFT_3$ dictionary to identify $k$ with the level and  $N$
with the rank of the gauge groups of the ABJ(M) theory.
All dissipative corrections enumerated in \cite{Rangamani:2009xk} are
seen to vanish, ensuring the uniqueness of the decomposition given in \eqref{T}.

Integrating the energy density, measured with respect to a static
geodesic observer $\rho_E$ yields
\begin{equation}
M= \int \rho _{E} \sqrt{-\gamma^{(0)}}~dxd\phi
=\frac{\alpha m}{K}\,,
\end{equation}
for the mass, in agreement with \eqref{confmass}, and independent
of the conformal frame (the choice of $\omega$).

We see from \eqref{bndymet}  that the boundary metric does not satisfy
Dirichlet boundary conditions. However for arbitrary variations of  the
parameters $A$ and $m~$ we find
that $\delta I[g]=0$ provided we set  $\alpha = \sqrt{1-A^2 \ell^2}$.  Our analysis therefore
points towards the possibility of generalizing the conditions under which
the conformal and holographic methods coincide for the mass computation
\cite{Hollands:2005wt,Papadimitriou:2005ii}.

\begin{figure}[tbp]
\centering
\includegraphics[width=0.45\textwidth]{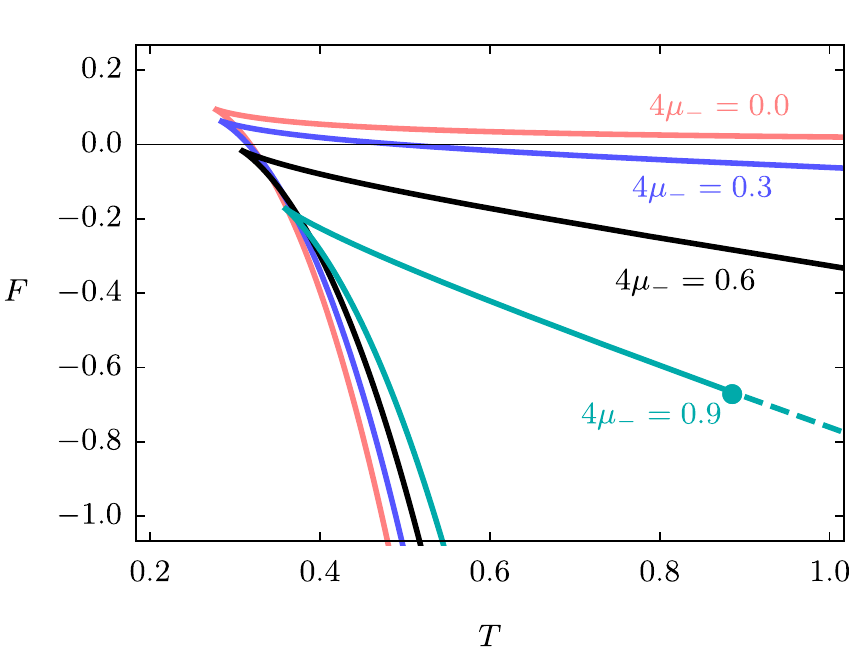}
\caption{\textbf{Free energy.}
The red curve is the Schwarzschild-AdS case, illustrating the
well-known Hawking--Page transition,  situated at a temperature
given by the intersection of the red curve with $F=0$. We do
not know of any such interpretation for all other curves with $\mu\neq 0$.
The upper parts of these curves do not continue to arbitrarily large $M$
but terminate at the boundary given in figure
\ref{fig:f1}; this is visible in the above plot only for $4\mu_- =0.9$.  }
\label{fig:FE}
\end{figure}

Finally, let us return to the computation of the action \eqref{Caction}.
We find
\begin{equation}
I = \frac{\beta}{2\alpha K} \left( m - 2mA^2\ell^2
- \frac{r_+^3}{\ell^2(1-A^2r_+^2)^2} \right)\,,
\label{Caction}
\end{equation}
using the time coordinate $\tau$. Some simple algebra then yields the
expected result $F=I/\beta=M-TS$
for the free energy, which we plot in figure \ref{fig:FE}.

Although similar in form, the behaviour of the free energy no longer indicates
the presence of a standard Hawking--Page transition \cite{Hawking:1982dh}.
As the string tension is fixed for the curves in the plot, no transition to
pure radiation (with zero tension) is possible. One may, however, speculate that
a transition to a different type of spacetime (for example {that of the
expanding spherical wave with an attached semi-infinite string of given
tension, similar to} \cite{Podolsky:2004bk}) may still be possible---such
an investigation, however, remains to be carried out.

We can also explore the isoperimetric ratio, or the ratio of volume to
areal radius: $\mathcal{R}=\left( \frac{3{V}}{\omega _{2}}\right)^{\frac{1}{3}}
\left(\frac{\omega _{2}}{{\cal A}}\right) ^{\frac{1}{2}}$ (recall $\omega_2=4\pi/K$
here). Using \eqref{entropy} and \eqref{vol} we find $\mathcal{R}\geq 1$,
indicating it satisfies the standard {\em reverse isoperimetric inequality}
\cite{Cvetic:2010jb}, not adding to the notable exceptions
\cite{Hennigar:2014cfa,Hennigar:2015cja,Brenna:2015pqa}.

Our full and consistent description of the thermodynamics of an
accelerating black hole reconciles discrepancies and conflicts that have
appeared in previous investigations of this system~\cite{Appels:2016uha,
Appels:2017xoe,Astorino:2016ybm}.  For example,  while a
set of thermodynamic variables for charged accelerating black holes
respecting the first law
was obtained \cite{Appels:2016uha,Appels:2017xoe,Gregory:2017ogk}
the computations employed an incorrectly normalized Killing
vector at infinity; furthermore  the resultant free energy is not consistent
with the standard Euclidean action calculation.
Alternate expressions for mass and temperature have been posited
\cite{Astorino:2016ybm}, with the tension of one deficit  held fixed to zero.
The other tension, while allowed to vary, was not included in the
first law, which was derived by assuming integrability of a scaling
of mass and temperature.  However no physical interpretation was given
either for this scaling or for why the energy content of the tension was
thermodynamically irrelevant.
Furthermore, the vacuum accelerating black hole
has an acceleration horizon, akin to a Rindler horizon, and the full
structure of the spacetime is that of two accelerating black holes in
two Rindler regions. Whether one should be considering a single
thermodynamic mass and first law with an additional horizon and black hole,
or whether, as suggested in \cite{Dutta:2005iy}, this should be considered
as a single system with a mass dipole is an open question.

We also found a decomposition of the dual stress energy tensor for the accelerating
black hole in terms of a perfect fluid plus conformal tensors.
We obtained a new ``universal'' coefficient, $\xi$ that is relevant for the
fluid/gravity correspondence in non-conformally flat manifolds.
It is natural to expect the existence of non-hydrodynamic corrections to
the energy momentum tensor for an even-dimensional CFT due to the
conformal anomalies. We have shown here that a similar picture arises
in the odd-dimensional case by explicitly constructing the relevant
non-hydrodynamic corrections necessary to provide a complete
holographic description of the system, c.f.\ \cite{deFreitas:2014lia}.

It would also be interesting to make a connection with
the weak coupling calculation of stress tensors in the presence of
conical deficits \cite{Dowker:1977zj}.
Future work will involve investigating accelerating black holes with rotation,
scalar fields~\cite{Anabalon:2009qt, Anabalon:2012ta}, and charge. The latter
system will be a challenge due to the asymptotic structure of the gauge field.

Since our computation is independent of the conformal frame,
we can compare to investigations of holographic C-metrics with an
acceleration horizon. For example, by choosing $\omega^2 =
(1-A^{2}\ell^{2}X)\alpha^{-2}$, we recover the form of the boundary metric
employed in \cite{Hubeny:2009kz}, and our coordinate transformation
\eqref{FGt} is now valid throughout $x\in[-1,1]$. However, if the condition
$ A^{2}\ell^{2} X <1$ is violated, then a black droplet/black funnel is present,
and we no longer have an equilibrium temperature for the system in general.
The boundary geometry corresponds to a black hole in a spatially compact
universe, and so there is no spatial asymptotic region as pointed out
in \cite{Hubeny:2009kz}. However, with the full conformal degree of
freedom present in our expression, we can easily remedy this shortcoming
by, for example, multiplying the $\omega$ above by $\frac{1}{\sqrt{1-x}}$,
giving an $AdS_{2}\times S^{1}$ asymptotic region at $x=1$ with the $AdS_{2}$
and $S^{1}$ radii being equal. If we multiply by $\frac{1}{\sqrt{1-x^2}}$ then
there are actually two $AdS_{2}\times S^{1}$ asymptotic regions at $x=\pm1$
and $\gamma^{(0)}_{ab}$ yields the geometry of a wormhole when there are no
horizons at the boundary.
The $AdS_{2}\times S^{1}$ asymptotic geometry is supersymmetric
and to our knowledge has been unnoticed so far in the literature.

\subsection*{Acknowledgements}
\vskip -0.3 cm
We thank Alex Buchel, Rob Myers, {Ji\v r\'i Podolsk\'y}, Mukund Rangamani and Kostas Skenderis for valuable discussions.
This work was
supported in part by the Natural Sciences and Engineering
Research Council of Canada, a Chilean FONDECYT [Grant No.\ 1181047 (AA),
1170279 (AA), 1161418 (AA), 3170035 (A.\"{O}.)],  CONYCT-RCUK [Newton-Picarte
Grants DPI20140053 (AA) and DPI20140115 (AA)], by the STFC
[consolidated grant ST/P000371/1 (RG)].
MA and AO would like to thank Perimeter Institute and University of
Waterloo for hospitality.
Research at Perimeter Institute is supported by the Government of Canada
through the Department of Innovation, Science and Economic
Development Canada and by the Province of Ontario through the Ministry of
Research, Innovation and Science.

\providecommand{\href}[2]{#2}
\begingroup\raggedright

\end{document}